\documentclass[aps, 
prl,
twocolumn,
superscriptaddress,
showpacs,preprintnumbers,amsmath,amssymb,
longbibliography,floatfix]{revtex4-2}
\usepackage[
colorlinks=true, 
pdfstartview=FitV, 
linkcolor=blue, 
citecolor=magenta, 
urlcolor=blue, 
bookmarks=true,
bookmarksnumbered=true,
pdftitle={},
pdfauthor={}
]{hyperref}

\usepackage[dvips]{graphicx}
\usepackage{bm,latexsym,amsmath,amssymb,amsfonts,mathrsfs,textcomp}
\usepackage{color}
\input{colordvi.tex}
\usepackage{comment}
\usepackage{simpler-wick}
\usepackage{tikz}
\usepackage[compat=1.1.0]{tikz-feynhand}
\usepackage[normalem]{ulem}

\newcommand\sect[1]{{\it #1.}---}

\newcommand{\diff}{\mathrm{d}} 
\newcommand{\rmi}{\mathrm{i}} 
\newcommand{\rme}{\mathrm{e}}

\newcommand{\ptc}[1]{{\bar{#1}}}

\newcommand{\pk}{\ptc{k}}
\newcommand{\pn}{\ptc{n}}

\newcommand\Lcal{\mathcal{L}}

\newcommand\Zbb{\mathbb{Z}}

\newcommand{\average}[1]{\langle#1\rangle}

\newcommand{\br}{\bm{r}}
\newcommand{\bk}{\bm{k}}
\newcommand{\bp}{\bm{p}}
\newcommand{\bq}{\bm{q}}

\newcommand{\bnab}{\bm{\nabla}}

\newcommand{\tilV}{\widetilde{V}}

\newcommand{\U}{\text{U}}

\newcommand{\with}{\quad\mathrm{with}\quad}

\newcommand{\eff}{\mathrm{eff}}

\newcommand{\micro}{\mathrm{micro}}

\begin{document}

\title{
Universal van der Waals Force Between Heavy Polarons in Superfluids
}

\author{Keisuke Fujii}
\email{fujii@thphys.uni-heidelberg.de}
\affiliation{Institut f\"{u}r Theoretische Physik, Universit\"{a}t Heidelberg, D-69120 Heidelberg, Germany}

\author{Masaru Hongo}
\email{hongo@phys.sc.niigata-u.ac.jp}
\affiliation{Department of Physics, Niigata University, Niigata 950-2181, Japan}
\affiliation{RIKEN iTHEMS, RIKEN, Wako 351-0198, Japan}

\author{Tilman Enss}
\email{enss@thphys.uni-heidelberg.de}
\affiliation{Institut f\"{u}r Theoretische Physik, Universit\"{a}t Heidelberg, D-69120 Heidelberg, Germany}

\begin{abstract}
 We investigate the long-range behavior of the induced Casimir interaction between two spinless heavy impurities, or polarons, in superfluid cold atomic gases.
 With the help of effective field theory (EFT) of a Galilean invariant superfluid, we show that the induced impurity-impurity potential at long distance universally shows a 
 relativistic van der Waals-like attraction ($\sim 1/r^7$) resulting from the exchange of two superfluid phonons.
 We also clarify finite temperature effects from the same two-phonon exchange process.
 The temperature $T$ introduces the additional length scale $c_s/T$ with the speed of sound $c_s$.
 Leading corrections at finite temperature scale as $T^6/r$ for distances $r \ll c_s/T$ smaller than the thermal length.  For larger distances the potential shows a nonrelativistic van der Waals behavior 
 ($\sim T/r^6$) instead of the relativistic one.
 Our EFT formulation applies not only to weakly coupled Bose or Fermi superfluids but also to those composed of strongly correlated unitary fermions with a weakly coupled impurity.
 The sound velocity controls the magnitude of the van der Waals potential, which we evaluate for the fermionic superfluid in the BCS-BEC crossover.
\end{abstract}

\maketitle

\sect{Introduction}%
The force between physical objects is one of the most elementary concepts in physics.
The development of quantum field theory demonstrates that the force is mediated by exchanging bosonic quanta such as pions in nuclear physics~\cite{Yukawa:1935xg} and 
gauge bosons in particle physics~\cite{Dirac1927,Yang-Mills1954,Utiyama1956}.
Since the long-range behavior is dominated by the lightest excitation of the system, the Nambu-Goldstone boson~\cite{Nambu:1961tp,Goldstone:1961eq,Goldstone:1962es} plays a central role in determining the long-range force in symmetry broken phases.
In fact, the pions --- the pseudo-Nambu-Goldstone bosons governing the long-range behavior of the nuclear force --- have established their place in the modern effective theory of nuclear forces (see, e.g., Refs.~\cite{Epelbaum:2008ga,Machleidt:2011zz,Hammer:2019poc}).

Recently impurities in Bosonic media, called Bose polarons, have been attracting much attention in cold atomic physics~\cite{Astrakharchik2004,Cucchietti2006,Palzer2009,Catani2012,Spethmann2012,Rath2013,Ardila2015,Levinsen2015,Grusdt2015,Hu2016,Jorgensen2016,Shchadilova2016,Yoshida2018,Camargo2018,Schmidt2018,Takahashi2019,yan2020bose,Drescher2020,skou2021,Massignan2021,Seetharam2021,Seetharam2021a}.
In particular, the two-impurity problem in a superfluid is an interesting playground, where the force mediated by collective excitations in the medium controls the impurity dynamics.
Similarly to the nuclear force, the induced interaction is often described by the Yukawa potential:
In fact, the single-Bogoliubov mode exchange has been shown to induce an attractive Yukawa potential ($\sim \rme^{- \sqrt{2} r/\xi}/r$) that falls off exponentially beyond the healing length $\xi$~\cite{Pethick2008,Nakano2016,Pascal2018,Camacho2018PRL,Camacho2018PRX};
in one dimension it leads to an attractive exponential potential ($\sim \rme^{-2r/\xi}$)~\cite{Klein2005,Recati2005,Dehkharghani2018}.
An exception are charged ionic impurities, where the bare atom-ion potential ($\sim 1/r^4$) dominates at large distances~\cite{Ding2022}.  
Note that induced interactions in a superfluid medium are attractive, and thereby easier to observe than the oscillatory Ruderman–Kittel–Kasuya–Yosida interaction in an ideal Fermi gas~\cite{Nishida2009,Macneill2011,Endo2013,Tilman2020}.

\begin{figure}[t]
 \centering
 \includegraphics[width=1.0\linewidth]{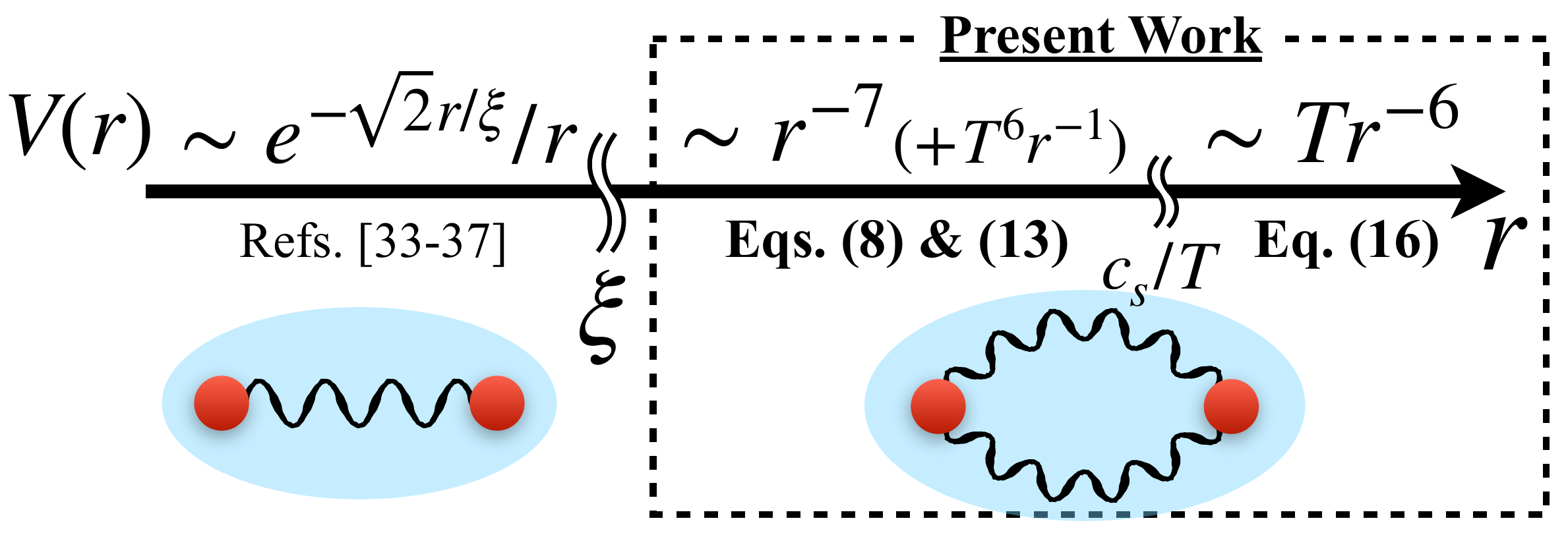}
 \caption{Schematic picture of the scaling regimes of the induced attractive Casimir interaction $V(r)$.
 } 
 \label{fig:regimes}
\end{figure}

In this Letter, we show that even for neutral impurities with a short-range potential, the exchange of two superfluid phonons generally leads to
a universal power-law induced interaction at a long distance that dominates over the Yukawa potential and becomes leading in the experimentally relevant regime $r\gtrsim\xi$.
The induced potential $V(r)$ is shown to be the relativistic van der Waals (Casimir) potential ($\sim 1/r^7$)~\cite{Casimir1948}
with a Coulomb correction ($\sim T^6/r$) at $\xi \ll r \ll c_s/T$ 
and the nonrelativistic van der Waals potential ($\sim T/r^6$) at 
$(\xi \ll) c_s/T \ll r$, where
$T$ and $c_s$ denote the temperature and the speed of sound (see Fig.~\ref{fig:regimes}). 
This extends previous results for the Casimir force in one   dimension~\cite{Schecter2014,Reichert2019a,Reichert2019b,Will2021} to higher dimension, and we present explicit results for the three-dimensional case.

Our formulation is based on a Galilean invariant superfluid EFT~\cite{Greiter:1989qb,Son:2005rv} with the assumption that the impurity is weakly coupled to the medium through $s$-wave contact interactions.  
While we assume weak impurity-medium coupling, the medium itself can be weakly or strongly coupled, including fermionic superfluids in the BCS-BEC crossover and, in particular, at unitarity \cite{zwerger2012}.
The magnitude of the potential is controlled by the sound velocity, which we can estimate from experimental data~\cite{Hoinka2017} for a fermionic superfluid in the BCS-BEC crossover.
While the power-law behavior arises at higher order in the gas parameter than the Yukawa potential, it becomes dominant in strongly correlated superfluids and toward the BCS regime. 

\sect{Superfluid EFT with impurities}%
\label{sec:EFT}%
We consider an attractive Fermi gas or a Bose gas weakly interacting with heavy impurities in the contact $s$-wave channel.
In the ground state these quantum gases form a superfluid with phonon excitations.
With the superfluid phonon field $\bar{\varphi}$ and impurity field $\Phi$, the low-energy superfluid EFT is described by the Lagrangian density
\begin{equation}
 \Lcal_{\eff}
  = p (\theta) 
  + \Phi^\dag \left( \rmi \partial_t + \frac{1}{2M} \bnab^2 \right) \Phi
  - g n (\theta) \Phi^\dag \Phi,
\end{equation}
where $p (\mu)$ and $n(\mu) =p^{\prime}(\mu)$ denote the medium pressure and number density as functions of the chemical potential $\mu$.
The first term describes the dynamics of the superfluid medium, the second is the impurity kinetic term, and the last term denotes the contact (zero-range) density-density interaction between impurity and medium.
The impurity-medium coupling constant can be expressed as $g = 2\pi a_{\mathrm{IM}}(\frac{1}{M} + \frac{1}{m})$ with the $s$-wave scattering length $a_{\mathrm{IM}}$ between the impurity (mass $M$) and medium particles (mass $m$).
By Galilean invariance of the superfluid medium the Lagrangian density depends on the phonon field only via the combination
$\theta \equiv \mu - \partial_t \bar{\varphi} - \frac{1}{2m} (\bnab \bar{\varphi})^2$~\cite{Greiter:1989qb,Son:2005rv}.

We expand $p(\theta)$ and $n(\theta)$ in gradients of the phonon field $\varphi \equiv\sqrt{\chi}\bar{\varphi}$,
rescaled by the compressibility $\chi = n^{\prime} (\mu)$, to obtain 
\begin{align}
 \Lcal_{\eff} 
 &= \frac{1}{2} (\partial_t \varphi)^2 
 - \frac{1}{2} c_s^2 (\bnab \varphi)^2
 + \Phi^\dag 
 \left( \rmi \partial_t + \frac{\bnab^2}{2M} - g \pn \right) 
 \Phi
 \nonumber \\
 &\quad
 + g
 \left[ 
 \sqrt{\chi} \partial_t \varphi + \frac{1}{2m} (\bnab \varphi)^2
 \right]
 \Phi^\dag \Phi 
 + \cdots ,
 \label{eq:eff-Lagrangian}
\end{align}
with speed of sound $c_s \equiv \sqrt{ \pn/(m \chi) }$ and average density $\pn = n (\mu)$.
In the Supplemental Material~\cite{SM} we derive this EFT explicitly from the microscopic theory of a weakly interacting Bose gas, but its form is a consequence of symmetry and holds also for strongly interacting superfluids~\cite{Son:2005rv, marini1998, Roberto2008, Schakel2011, Klimin2014}.
The first line of Eq.~\eqref{eq:eff-Lagrangian} leads to the phonon propagator
\begin{equation}
 \rmi G (p) = \frac{\rmi}{(p^0)^2 - E_{\bp}^2 + \rmi \epsilon} 
  \with
  E_{\bp} \equiv c_s |\bp|,
  \label{eq:phonon-propagator}
\end{equation} 
while the second line describes the interaction with the impurities.
We emphasize that Galilean invariance is crucial to identify the two-body coupling between the impurity and the phonons (see Ref.~\cite{Schecter2014} for the same result in one-dimensional systems from a slightly different perspective).
In the following, we drop the ellipsis part in Eq.~\eqref{eq:eff-Lagrangian} as a Galilean invariant truncation.
As elaborated in~Ref.~\cite{Son:2005rv}, the higher-order phonon terms are highly suppressed at low energy due to the derivative interaction, and we can safely neglect them.%
\footnote{The ellipsis part contains also a term $g\frac{\chi^{\prime}}{2\chi}(\partial_t\varphi)^2\Phi^{\dagger}\Phi$ of second order in $\varphi$ with $\chi^{\prime}=\chi^{\prime}(\mu)$. The two-phonon exchange from this term gives rise to a power-law potential ($\sim r^{-7}$) at zero temperature~\cite{Pavlov2018,Pavlov2019} that is parametrically smaller than our result~\eqref{eq:V1} and \eqref{eq:V3} by a factor of $mc_s^2\chi^{\prime}/\chi\sim (c_s/k_F)^2$. At finite temperature it yields an exponential decay ($\sim r^{-2}e^{-4\pi Tr/c_s}$) at longer distances $r\gg c_s/T$ because the interaction vertex of the time-derivative coupling is proportional to the frequency and has no contribution from the Matsubara zero mode.}

\sect{Potential from the exchange of two superfluid phonons}%
\label{sec:potential}%
In this Letter, we focus on the long-range behavior of the induced interaction between two heavy impurities, which allows us to treat them as test particles fixed at a certain distance $r$.
Assuming that the impurity-medium coupling $g$ is small, we will evaluate the leading-order induced potential from Feynman diagrams of phonon exchange~\cite{Schecter2014}.
The impurity kinetic term involving $\Phi$ in the first line of Eq.~\eqref{eq:eff-Lagrangian} does not affect the potential at leading order and can be neglected for heavy impurities.

The exchange of a single Bogoliubov mode produces the Yukawa potential $e^{-\sqrt2r/\xi}/r$ \cite{Pethick2008,Nakano2016,Pascal2018,Camacho2018PRL,Camacho2018PRX} that arises from the nonlinearity of the Bogoliubov dispersion.  In the low-energy regime $r\gg\xi$, however, only the linear phonon branch remains and the Yukawa potential vanishes in the limit $\xi/r\to0$. This behavior is directly obtained within our low-energy EFT: combining two interaction vertices $g \sqrt{\chi} \Phi^\dag \Phi \partial_t \varphi$ from Eq.~\eqref{eq:eff-Lagrangian} yields a contribution to the induced potential at leading order $g^2$. The static potential induced by this exchange of a single static phonon carrying $k=(0,\bk)$ vanishes because the interaction vertex is proportional to the frequency $k^0=0$.

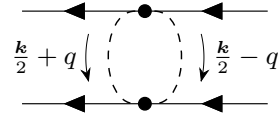
\begin{figure}[t]
 \centering
 \begin{equation*}
  \scalebox{1.1}{
   \begin{tikzpicture}[baseline=(o.base)]
    \begin{feynhand}
     \vertex (o) at (0,-0.1) {};   
     \vertex (a) at (1.6,0.56) {};
     \vertex (b) at (1.6,-0.56) {};
     \vertex (v1) at (0,0.56) {};
     \vertex [dot] at (0,0.56) {}; 
     \vertex (c) at (-1.6,0.56) {};
     \vertex (d) at (-1.6,-0.56) {};
     \vertex (v2) at (0,-0.56) {};
     \vertex [dot] at (0,-0.56) {}; 
     \propag [with arrow=0.22, with arrow =0.78] (a) to (c);
     \propag [with arrow=0.22, with arrow =0.78] (b) to (d);
     \propag [sca, mom'={[xshift=-0.6mm, arrow shorten=0.35] $\frac{\bk}{2} + q$}] (v1) to [out=180,in=180, looseness=1] (v2);
     \propag [sca, mom={[xshift=0.6mm, arrow shorten=0.35] $\frac{\bk}{2} - q$}] (v1) to [out=0,in=0, looseness=1] (v2);
    \end{feynhand}
   \end{tikzpicture}}
 \end{equation*}
 \caption{The exchange of two superfluid phonons (dashed lines) gives rise to the induced potential between two impurities (amputated solid lines).} 
 \label{fig:two-phonon-exchange}
\end{figure}

On the other hand, the second interaction vertex $g\frac{(\bnab \varphi)^2}{2m}\Phi^\dagger \Phi$ leads to a two-phonon exchange process at the same order $g^2$ as illustrated in Fig.~\ref{fig:two-phonon-exchange}.
Although it appears at higher order in the inverse compressibility $\chi^{-1}$ or of the BEC gas parameter, we find that it gives the leading result at long distance: a power law that dominates over the exponentially suppressed Yukawa potential.
The two-phonon exchange leads to the induced potential in Fourier space as
\begin{align}
 - \rmi \tilV (\bk)
 = - \frac{g^2}{2 m^2} 
 & \int \frac{\diff^4 q}{(2\pi)^4}
 \left( \frac{\bk^2}{4} - \bq^2 \right)^2
 \nonumber \\
 &\times
  \rmi G \left( \frac{k}{2} + q \right) 
  \rmi G \left( \frac{k}{2} - q \right), 
  \label{eq:V1}
\end{align}
with $k = (0,\bk)$ and $q = (q^0,\bq)$.
Using the phonon propagator [Eq.~\eqref{eq:phonon-propagator}] and the dimensional regularization, we evaluate the $q$ integral in Eq.~\eqref{eq:V1} as~\cite{SM}
\begin{equation}
 \tilV (\bk) = 
  \frac{g^2 \bk^4}{32 \pi^2 m^2 c_s^{3}}
  \left(
   \frac{43}{240} \log \frac{\bk^2}{\lambda^2}
   - \frac{8261}{1440}
  \right),
  \label{eq:V2}
\end{equation}
where we employed the modified minimal subtraction ($\overline{\mathrm{MS}}$) scheme with the renormalization scale $\lambda$ (see, e.g., Ref.~\cite{Peskin:1995ev})%
\footnote{\label{footnote:regularization}
The second term in Eq.~\eqref{eq:V2} without the logarithm depends on the regularization method and does not affect the long-range behavior of the induced potential.}.

To obtain the induced potential $V(\br_1 - \br_2)$ between two impurities at positions $\br_1$ and $\br_2$, one needs to perform the Fourier transform of $\tilV (\bk)$, which is clearly UV divergent.
One can correctly read off the finite potential by introducing an appropriate convergence factor as
\begin{align}
 V(\br_1-\br_2)= \lim_{\epsilon \to 0^{+}}\int \frac{\diff^3k}{(2\pi)^3}  \rme^{\rmi\bk\cdot(\bm{r}_1-\bm{r}_2)- \epsilon|\bk|} \tilV(\bk).
 \label{eq:potential-Fourier}
\end{align}
With the help of the formula~\cite{berestetskii1982quantum}
\begin{equation}
\lim_{\epsilon \to 0^{+}}\int \frac{\diff^3 k}{(2\pi)^3}|\bk|^{\nu}
 \rme^{\rmi\bk\cdot\bm{r}-\epsilon|\bk|}
= -\frac{\Gamma(\nu+2)\sin(\nu\pi/2)}{2\pi^2 |\bm{r}|^{\nu+3}}
\label{eq:regulator-method}
\end{equation}
and its derivative with respect to $\nu$,
we perform the Fourier transform in Eq.~\eqref{eq:potential-Fourier} to
obtain
\begin{align}
 V (\br_1 - \br_2)
 = - \frac{43g^2}{128 \pi^3 m^2 c_s^{3}} 
 \frac{1}{|\br_1 - \br_2|^7}.
 \label{eq:V3}
\end{align}
We thus find that the long-range behavior of the impurity potential in the superfluid is not given by the Yukawa potential but by the relativistic version of the van der Waals potential~\cite{berestetskii1982quantum}.

\sect{Finite-temperature effect}%
\label{sec:coefficient}%
One can investigate the effect of finite temperature on the induced potential with the help of the Matsubara formalism~\cite{Matsubara,AGD}.
For that purpose, we need to replace the phonon propagator [Eq.~\eqref{eq:phonon-propagator}] with 
\begin{equation}
 \Delta (\rmi \omega_n, \bk) = \frac{1}{\omega_n^2 + E_{\bk}^2}
  \with 
  \omega_n \equiv 2 \pi n T,
\end{equation}
where we introduced the temperature $T$ and the bosonic Matsubara frequency $\omega_n$ with $n \in \Zbb$.
Then, the impurity potential at $T > 0$ is given by
\begin{align}
 \tilV_{T} (\bk) 
 &= -\frac{g^2}{2 m^2} 
 T \sum_{n= -\infty}^{\infty}
 \int \frac{\diff^3 q}{(2\pi)^3} 
 \left( \frac{\bk^2}{4} - \bq^2 \right)^2
 \nonumber \\
 &\quad \times
 \Delta \left( \rmi \omega_n, \frac{\bk}{2} + \bq \right)
 \Delta \left( - \rmi \omega_n, \frac{\bk}{2} - \bq \right).
 \label{eq:potential-finite-T}
\end{align}
Since the temperature introduces the additional length scale 
$c_s/T$ in our problem, 
there emerge two subregimes for the potential $V_T(r)$ at finite temperature, for interparticle distances shorter or longer compared with $c_s/T$.

First, at intermediate distances $\xi \ll r\ll c_s/T$, which covers the whole long-distance regime at zero temperature, the potential acquires an additional finite-temperature correction 
as $\tilV_T(\bk) = \tilV (\bk) + \Delta \tilV_T (\bk)$.
Computing the Matsubara sum in Eq.~\eqref{eq:potential-finite-T} we find the finite-temperature correction $\Delta \tilV_T(\bk)$ as
\begin{align}
 \!\!\! 
 \Delta \tilV_T (\bk)
 &=-\frac{g^2}{2 m^2} 
 \int \frac{\diff^3 q}{(2\pi)^3} 
 \left( \frac{\bk^2}{4} - \bq^2 \right)^2
 \frac{1}{2E_+ E_-}
 \nonumber \\
 &~\times
 \left[
 \frac{f (E_+) + f (E_-)}{E_+ + E_-} 
 - \frac{f (E_+) - f (E_-)}{E_+ - E_-} 
 \right],
\end{align}
where we introduced the Bose distribution
$f (E) = 1/(\rme^{\beta E} - 1)$ and $E_{\pm} \equiv E_{\bk/2 \pm \bq}$.
The low-temperature expansion allows us to obtain the analytic expression~\cite{SM}
\begin{equation}
 \Delta \tilV_T (\bk)
 \simeq -\frac{g^2}{32 \pi^2 m^2 c_s^3}
 \frac{128 \pi^6}{135 c_s^6} \frac{T^6}{\bk^2} 
 ~~\mathrm{at}~~
 T \ll c_s |\bk| ,
 \label{eq:kspace-potential-T}
\end{equation}
where we omit the constant term.
Therefore, we find a Coulomb potential as the low-temperature correction 
\begin{equation}
 \Delta \tilV_T (\br_1 - \br_2) 
  \simeq - \frac{g^2}{128 \pi^3 m^2 c_s^3}
 \frac{128 \pi^6}{135 c_s^6} \frac{T^6}{|\br_1 - \br_2|},
 \label{eq:potential-T}
\end{equation}
which is suppressed by a factor $(T|\br_1 - \br_2|/c_s)^6 \ll 1$ compared to the 
relativistic van der Waals potential \eqref{eq:V3}.

At longer distances $r\gg c_s/T$, the Matsubara zero mode 
$\omega_{n}= 0$ gives the dominant contribution in Eq.~\eqref{eq:potential-finite-T},
and the full $\tilV_T(\bk)$ is approximated as
\begin{align}
\tilV_T(\bk) \simeq
 -\frac{g^2}{2 m^2} 
 T &\int \frac{\diff^3 q}{(2\pi)^3} 
 \left( \frac{\bk^2}{4} - \bq^2 \right)^2
 \nonumber \\
 &\times
 \Delta \left(0, \frac{\bk}{2} + \bq \right)
 \Delta \left(0, \frac{\bk}{2} - \bq \right).
\end{align}
With the use of the dimensional regularization
it is again straightforward to perform this integral
as~\cite{SM}
\begin{equation}
 \tilV_T(\bk)
 \simeq
 -\frac{g^2}{64 m^2c_s^4}T|\bk|^3
 ~~\mathrm{at}~~
 c_s |\bk| \ll T
 .
 \label{eq:potential-k-T-long}
\end{equation}
Using Eq.~\eqref{eq:regulator-method} as before, we find the induced potential as
\begin{equation}
 V_T(\br_1 - \br_2)
 =-\frac{3g^2}{16 \pi^2 m^2 c_s^4}\frac{T}{|\bm{r}_1-\bm{r}_2|^6},
 \label{eq:potential-r-T-long}
\end{equation}
which is the familiar nonrelativistic van der Waals potential proportional to $1/r^6$.

In short, we find the induced potential for two subregimes separated by the temperature length $c_s/T$;
it acquires the finite Coulomb-type correction at intermediate distances $\xi \ll r \ll c_s/T$, while it approaches asymptotically the nonrelativistic van der Waals potential at longer distances $c_s/T \ll r$ (see Fig.~\ref{fig:regimes}).
Note that the induced potential still exhibits a power-law decay rather than the Yukawa potential in both regimes.
This is because the superfluid phonon remains exactly gapless even at finite temperature when $\U(1)$ symmetry is spontaneously broken in three dimensions.

It is worth emphasizing that the long-range van der Waals behavior (with the Coulomb correction at $T>0$) is universal, i.e.,
independent of the detailed microscopic parameters of the model.
The result follows once we assume that the Galilean invariant medium is in the superfluid phase supporting the gapless phonon, and the impurity is weakly coupled via an $s$-wave contact interaction.
Thus, our results [Eqs.~\eqref{eq:V3}, \eqref{eq:potential-T} and \eqref{eq:potential-r-T-long}] are valid in the entire BCS-BEC crossover, including the strongly correlated unitary Fermi gas regime.

\sect{Magnitude of the potential in the BCS-BEC crossover}%
\label{sec:magnitude}%
While the power-law exponent of the van der Waals potential is universal, the magnitude of the potential depends on the medium properties through the speed of sound $c_s$.  
This input parameter for our EFT is determined, e.g., from experimental data for fermionic superfluids~\cite{Hoinka2017} or from microscopic theoretical calculations~\cite{marini1998,Roberto2008,Schakel2011}.
Focusing on a fermionic superfluid in the BCS-BEC crossover, we shall evaluate the magnitude of the van der Waals potential in comparison to the Yukawa potential.

\begin{figure}[t]
 \centering
 \includegraphics[width=0.95\linewidth]{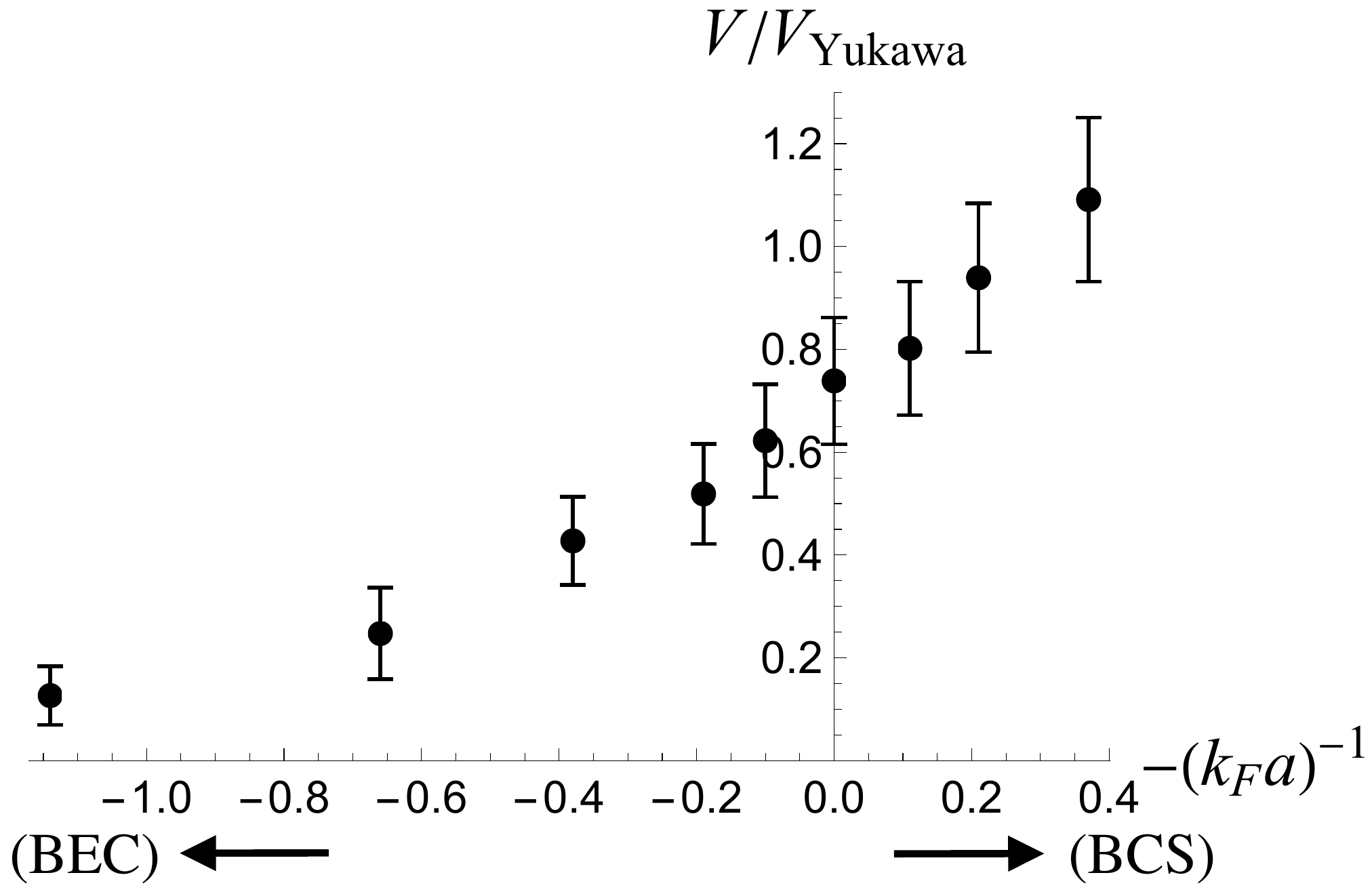}
 \caption{
 Strength of the van der Waals interaction $V(r)$ in Eq.~\eqref{eq:V3} compared to the Yukawa potential $V_{\textrm{Yukawa}}(r)=-g^2 m\bar{n} \rme^{-\sqrt{2}r/\xi}/(2\pi r)$ at fixed $r=8\xi$.
 Our EFT result is evaluated using the experimental data~\cite{Hoinka2017} for the speed of sound $c_s/v_F$ as a function of the interaction parameter $-(k_Fa)^{-1}$ with $s$-wave scattering length $a$ and Fermi momentum $k_F$.  The importance of the van der Waals potential grows from the weakly coupled molecular BEC (left) to dominate in the unitary and BCS regimes (right).}
 \label{fig:magnitude}
\end{figure}

Using the experimental reference data~\cite{Hoinka2017}, we demonstrate the ratio of our result [Eq.~\eqref{eq:V3}] to the Yukawa potential in Fig.~\ref{fig:magnitude}.
The Yukawa potential from the exchange of a single Bogoliubov mode was obtained in Ref.~\cite{Pascal2018} as $V_{\textrm{Yukawa}}(r)=-g^2 m \bar{n}  \rme^{-\sqrt{2}r/\xi}/(2\pi r)$
with the healing length $\xi\equiv 1/(\sqrt{2}mc_s)$
\footnote{\label{footnote:coeff-Yukawa}
The Yukawa potential $V_{\textrm{Yukawa}}(r)$ with $g=2\pi a_{\textrm{IM}}/m$ looks different from the original one in Ref.~\cite{Pascal2018} by a factor of $1/2$ because the boson density is half the fermion density, $\bar{n}/2$ in our notation.
}.
The result is shown as a function of the dimensionless medium interaction parameter $-(k_F a)^{-1}$ with Fermi momentum $k_F$ and $s$-wave scattering length $a$ of the medium fermions
\footnote{
\label{footnote:Yukawa}
We express the result for the Yukawa interaction~\cite{Pascal2018} in terms of the speed of sound $c_s$ and then apply it to the BCS-BEC crossover using the input data for $c_s/v_F$ as a function of $-(k_F a)^{-1}$}.
One sees that the contribution from the van der Waals potential becomes relatively larger 
when $-(k_Fa)^{-1}$ increases, and it dominates toward the BCS side. 

\sect{Discussion and outlook}%
\label{sec:Discussion}%
In this Letter, we have clarified the universal long-range behavior of the potential between impurities based on Galilean invariant superfluid EFT.
We find that the exchange of two superfluid phonons leads to the relativistic van der Waals potential $V (\bm{r}) \sim 1/r^7$ at zero temperature.
We also find that at finite temperature $T>0$ it leads to the nonrelativistic van der Waals potential $V (\bm{r}) \sim T/r^6$ at larger distances $r\gg c_s/T$,
while the potential acquires a Coulomb-type correction $\Delta V_T (r) \sim T^6/r$ at intermediate distances $\xi \ll r \ll c_s/T$.
The result is universal since the EFT only relies on two assumptions:
(i) the medium is a Galilean invariant superfluid, and (ii) the impurity is weakly coupled to the medium through $s$-wave contact interactions.

This power-law potential always dominates over the Yukawa potential at large distance $r\gg\xi$.  
In addition, we have shown in Fig.~\ref{fig:magnitude} that even at fixed distance $r/\xi$, the relative importance of the van der Waals potential increases monotonically
with the interaction strength $-(k_Fa)^{-1}$, indicating that it dominates already at shorter distances in the strongly coupled and BCS regimes of atomic gases.
Our EFT cannot capture the behavior at distances shorter than the healing length $\xi$, where nonlinear terms in the phonon dispersion appear.
Since $\xi$ decreases from large distances on the weakly coupled BEC side toward values as short as the particle spacing in the strongly coupled unitary gas~\cite{Engelbrecht1997},
both superfluidity and the van der Waals potential are more robust at strong coupling.
It is therefore highly desirable to further investigate the properties of impurities in fermionic superfluids~\cite{Yi2015,Nishida2015,Laurent2017,Pierce2019,Castin2020,Bigue2022,Wang2022}.

The experimental observation of this Casimir interaction should be feasible with present technology using ultracold quantum gases.  
Each impurity experiences a mean-field energy shift $E_\text{pol}= O(g)$ and in addition the smaller Casimir shift $V(r)= O(g^2)$ due to the presence of a second impurity.  
The effect of the power-law van der Waals scaling, as opposed to exponential Yukawa scaling, is most pronounced at distances $r\sim 5\dotsc 10\,\mu$m a few times larger than the healing length $\xi\lesssim1\,\mu$m.  
Even a small Casimir shift can be detected by Ramsey interferometry: first, two fermionic impurities in identical spin states experience the mean-field shift but no $s$-wave channel contribution in the scattering under the induced interaction.
Second, two fermionic impurities in distinct spin states experience both the mean-field shifts and the full induced interaction.  When both time-evolved states are superimposed, even a small energy shift from the induced interaction will result in observable interference fringes.
Alternatively, the induced interaction can lead to an observable shift in the oscillation frequency of two impurities confined to separate microtraps in a recently proposed experimental setup~\cite{Ding2022}.

While we focused on weak interaction between the impurity and the medium perturbatively, possible nonperturbative effects are worth further investigation~\cite{Will2021, Ding2022}.
For example, the strong attractive interaction between an impurity and medium particles may lead to the formation of bound states~\cite{Drescher2020,Schmidt2022}.
Furthermore, even for a weakly interacting BEC, as the impurity-medium coupling $g$ approaches the bound-state threshold, the induced potential could lead to an Efimov attraction that can bind two impurities~\cite{Pascal2018}.
It is interesting to investigate the universality of such bound states at long distances governed by superfluid phonons.
Besides, it is worth extending our formulation to more general cases,
e.g., systems with a dipolar interaction between the medium and impurity, or distinct symmetry broken phases of a spinor BEC. 
These extensions may lead to different universal behavior for the impurity problem;
we leave these for future work.

\smallskip
\begin{acknowledgments}
The authors thank G.~Bighin, M.~Drescher, S.~Endo, T.~Hatsuda, Y.~Hidaka, P.~Naidon, and Y.~Nishida for useful discussions.
This work is supported by the Deutsche Forschungsgemeinschaft (DFG, German Research Foundation), Project-ID 273811115 (SFB1225 ISOQUANT) and under Germany’s Excellence Strategy EXC2181/1-390900948 (the Heidelberg STRUCTURES Excellence Cluster).
M.H. was partially supported by the U.S. Department of Energy, Office of Science, Office of Nuclear Physics under Award No.~DE-FG0201ER41195 and JSPS KAKENHI Grant-in-Aid for Research Activity Start-up 22K20369.
This work is partially supported by the RIKEN iTHEMS Program (in particular iTHEMS Non-Equilibrium Working group and iTHEMS Mathematical Physics Working group).
\end{acknowledgments}

\bibliography{polaron-force}

\appendix
\pagebreak
\widetext
\begin{center}
\textbf{\large Supplemental Materials:\\
Universal van der Waals force between heavy polarons in superfluids}
\end{center}
\setcounter{equation}{0}
\setcounter{figure}{0}
\setcounter{table}{0}
\setcounter{page}{1}
\newcounter{supplementeqcountar}
\newcommand\suppsect[1]{{\it #1.}---}
\renewcommand{\theequation}{S\arabic{equation}}
\renewcommand{\thefigure}{S\arabic{figure}}

\section{
Derivation of Eq.~(2) for the weakly intearacting Bose gas medium
}
\label{sec:mean-field-derivation}

We provide a derivation of our effective Lagrangian 
density from a microscopic theory 
for a weakly interacting Bose gas medium
relying on the derivative expansion.
Let us start with the following Lagrangian density,
\begin{align}
 \Lcal_{\micro}
 =\Phi^{\dagger}\biggl(\rmi \partial_t+\frac{1}{2M}\bnab^2\biggr)\Phi
 +\psi^{\dagger}\biggl(\rmi\partial_t+\frac{1}{2m}\bnab^2+\mu\biggr)\psi
 -\frac{g_{BB}}{2}\bigl(\psi^{\dagger}\psi\bigr)^2
 -g\psi^{\dagger}\psi\Phi^{\dagger}\Phi,
\end{align}
with the bosonic medium field $\psi$ and the impurity field $\Phi$.
For small $g_{BB}$, the ground state expectation value of $\psi$ is classically determined so as to minimize the potential for the medium
$-\mu\bigl(\psi^{\dagger}\psi\bigr)+\frac{g_{BB}}{2}\bigl(\psi^{\dagger}\psi\bigr)^2$. 
As a result, the ground state supports a non-vanishing Bose-Einstein condensate (BEC) $\average{|\psi|} = \sqrt{\mu/g_{BB}}$, leading to a spontaneous U(1) symmetry breaking.

To describe fluctuations above the U(1) symmetry broken ground state,
it is useful to parameterize the field $\psi$ as
\begin{align}
\psi=\sqrt{\bar{n}+\delta n}e^{i\bar{\varphi}}
\with
\bar{n}=\mu/g_{BB},
\end{align}
where $\delta n$ and $\bar{\varphi}$ represent amplitude and phase fluctuations, respectively.
Substituting this decomposition into $\Lcal_{\micro}$ leads to
\begin{align}
 \Lcal_{\micro}
 = \Phi^{\dagger}\biggl(\rmi \partial_t+\frac{1}{2M}\bnab^2-g\bar{n}\biggr)\Phi
 -\frac{\bar{n}}{2m}\bigl(\bnab\bar{\varphi}\bigr)^2
 -\frac{\bigl(\bnab\delta n\bigr)^2}{8m (\bar{n}+\delta n)}
 -\frac{g_{BB}}{2}\delta n^2
 -\delta n
 \biggl(
    g\Phi^{\dagger}\Phi
    +\partial_t\bar{\varphi}
    +\frac{\bigl(\bnab\bar{\varphi}\bigr)^2}{2m}
\biggr),
\end{align}
where we dropped constant and total derivative terms.
Assuming $\delta n\ll \bar{n}$ and truncating the Lagrangian density up to second order with respect to $\delta n$, we can derive the equation of motion for $\delta n$ and find its solution formally as
\begin{align}
 \biggl(g_{BB}-\frac{\bnab^2}{4m\bar{n}}\biggr)\delta n
 =-\biggl(
    g\Phi^{\dagger}\Phi
    +\partial_t\bar{\varphi}
    +\frac{\bigl(\bnab\bar{\varphi}\bigr)^2}{2m}
\biggr)
\quad \Rightarrow \quad
\delta n=-\frac{1}{g_{BB}\bigl(1-\frac{(\xi\bnab)^2}{2}\bigr)}\biggl(
    g\Phi^{\dagger}\Phi
    +\partial_t\bar{\varphi}
    +\frac{\bigl(\bnab\bar{\varphi}\bigr)^2}{2m}
\biggr),
\end{align}
where we introduced the healing length $\xi\equiv1/\sqrt{2m\bar{n}g_{BB}}$.
The effective Lagrangian $\Lcal_{\eff}$ is derived from the substitution of this solution into $\Lcal_{\micro}$ and is obtained as
\begin{align}
\begin{split}
 \Lcal_{\eff}
 &\simeq
 \Phi^{\dagger}\biggl(\rmi \partial_t+\frac{1}{2M}\bnab^2-g\bar{n}\biggr)\Phi
 -\frac{\bar{n}}{2m}\bigl(\bnab\bar{\varphi}\bigr)^2 \\
 &\quad
 +\frac{1}{2}
 \biggl(
    g\Phi^{\dagger}\Phi
    +\partial_t\bar{\varphi}
    +\frac{\bigl(\bnab\bar{\varphi}\bigr)^2}{2m}
\biggr)
\frac{1}{g_{BB}\bigl(1-\frac{(\xi\bnab)^2}{2}\bigr)}
\biggl(
    g\Phi^{\dagger}\Phi
    +\partial_t\bar{\varphi}
    +\frac{\bigl(\bnab\bar{\varphi}\bigr)^2}{2m}
\biggr).
\label{eq:microL-after-integration}
\end{split}
\end{align}

We focus on the long-distance physics $r\gg \xi$ 
(or $\bnab \ll \xi^{-1}$) and expand the second line of Eq.~\eqref{eq:microL-after-integration} with the use of the 
derivative expansion $1-\frac{(\xi\bnab)^2}{2}\simeq 1$.
Since the compressibility of the weakly interacting Bose gas is given by $\chi\equiv n^{\prime}(\mu)=1/g_{BB}$ in the mean-field approximation,
the Lagrangian density turns out to be
\begin{align}
 \Lcal_{\eff}
 &\simeq
 \Phi^{\dagger}\biggl(\rmi \partial_t+\frac{1}{2M}\bnab^2-g\bar{n}\biggr)\Phi
 +\frac{1}{2}\bigl(\partial_t\varphi\bigr)^2
 -\frac{1}{2}c_s^2\bigl(\bnab\bar{\varphi}\bigr)^2
 +g\Phi^{\dagger}\Phi
  \biggl(
    \sqrt{\chi}\partial_t\varphi
    +\frac{\bigl(\bnab\varphi\bigr)^2}{2m}
\biggr)+\cdots,
\label{eq:Lag-derivation}
\end{align}
where we used the rescaled field $\varphi=\sqrt{\chi}\bar{\varphi}$.
Here, we also introduced the speed of sound for the weakly interacting Bose gas as $c_s^2=g_{BB}\bar{n}/m$.
Equation~\eqref{eq:Lag-derivation} coincides with Eq.~\eqref{eq:eff-Lagrangian} in the main text
and accomplishes the microscopic derivation of our superfluid EFT for the weakly interacting Bose gas.

\section{Evaluation of two-phonon exchange potential at $T=0$}
\label{sec:two-loop-evaluation-T=0}

\begin{figure}[b]
 \centering
 \begin{equation*}
  \scalebox{1.0}{
  \begin{tikzpicture}[baseline=(o.base)]
   \begin{feynhand}
    \vertex (o) at (0,-0.1) {};
    \vertex (a) at (1.0,0) {};
    \vertex [dot] at (1.0,0) {}; 
    \vertex (b) at (-1.0,0) {};
    \vertex [dot] at (-1.0,0) {}; 
    \propag [sca, mom'={[arrow shorten=0.25] $p$}] (a) to (b);
   \end{feynhand}
  \end{tikzpicture}}
  = \rmi G (p) 
  \qquad 
  \scalebox{1.0}{
  \begin{tikzpicture}[baseline=(o.base)]
   \begin{feynhand}
    \vertex (o) at (0,-0.1) {};
    \vertex (f1) at (1.5,0) {};
    \vertex (f2) at (-1.5,0) {};
    \vertex (v) [dot] at (0,0) {};
    \vertex (d) at (0,1.2) {};
    \propag [fer] (f1) to (v);
    \propag [fer] (v) to (f2);
    \propag [sca, mom'={[arrow shorten=0.25] $p$}] (d) to (v);
   \end{feynhand}
  \end{tikzpicture}}
  = g \sqrt{\chi} p^0 
  \qquad 
  \scalebox{1.0}{
  \begin{tikzpicture}[baseline=(o.base)]
   \begin{feynhand}
     \vertex (o) at (0,-0.1) {};
    \vertex (f1) at (1.5,0) {};
     \vertex (f2) at (-1.5,0) {};
    \vertex (v) [dot] at (0,0) {};
     \vertex (d1) at (0.5,1.2) {};
     \vertex (d2) at (-0.5,1.2) {};
     \node at (0.7,0.55) {$q$};
     \node at (-0.7,0.55) {$p$};
     \propag [fer] (f1) to (v);
     \propag [fer] (v) to (f2);
     \propag [sca, mom={[arrow shorten=0.25]}] (d1) to (v);
     \propag [sca, mom'={[arrow shorten=0.25]}] (d2) to (v);
   \end{feynhand}
  \end{tikzpicture}}
   = - \frac{\rmi g}{2m} \bp \cdot \bq . 
 \end{equation*}
 \caption{The Feynman rule for the superfluid phonon and impurity.} 
 \label{fig:Feynmann-T=0}
\end{figure}
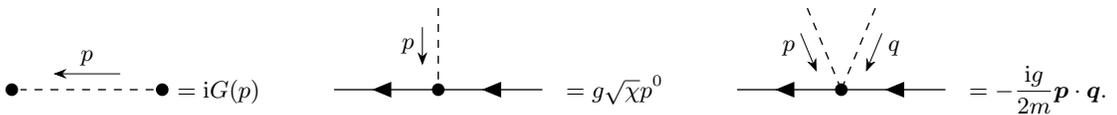

The leading-order effective Lagrangian density of the Galilean invariant superfluid coupled to impurities is given by 
\begin{align}
 \Lcal_{\eff} 
 =& \frac{1}{2} (\partial_t \varphi)^2 
 - \frac{1}{2} c_s^2 (\bnab \varphi)^2
 + \Phi^\dag 
 \left( \rmi \partial_t + \frac{1}{2M} \bnab^2 - g \pn \right) 
 \Phi
 + g
 \left[ 
 \sqrt{\chi} \partial_t \varphi + \frac{1}{2m} (\bnab \varphi)^2
 \right]
 \Phi^\dag \Phi ,
\end{align}
from which we find the phonon propagator 
\begin{equation}
 \rmi G (p) = 
  \frac{\rmi}{(p^0)^2 - E_{\bp}^2 + \rmi \epsilon}
  \with
  E_{\bp} \equiv c_s |\bp|. 
\end{equation}
and the interaction vertices (see Fig.~\ref{fig:Feynmann-T=0} for the Feynman rule).

We are interested in the potential between the two impurities.
Thus, one-superfluid phonon exchange contribution vanishes because it is proportional to the frequency. 
The leading-order impurity potential at $O(g^2)$ is given by 
the following two-superfluid phonon exchange 
\begin{equation}
 - \rmi \tilV (\bk)
  = \scalebox{1.0}{
   \begin{tikzpicture}[baseline=(o.base)]
    \begin{feynhand}
     \vertex (o) at (0,-0.1) {};   
     \vertex (a) at (1.75,0.8) {};
     \vertex (b) at (1.75,-0.8) {};
     \vertex (v1) at (0,0.8) {};
     \vertex [dot] at (0,0.8) {}; 
     \vertex (c) at (-1.75,0.8) {};
     \vertex (d) at (-1.75,-0.8) {};
     \vertex (v2) at (0,-0.8) {};
     \vertex [dot] at (0,-0.8) {}; 
     \propag [with arrow=0.22, with arrow =0.78] (a) to (c);
     \propag [with arrow=0.22, with arrow =0.78] (b) to (d);
     \propag [sca, mom'={[arrow shorten=0.35] $\frac{\bk}{2} + q$}] (v1) to [out=180,in=180, looseness=1] (v2);
     \propag [sca, mom={[arrow shorten=0.35] $\frac{\bk}{2} - q$}] (v1) to [out=0,in=0, looseness=1] (v2);
    \end{feynhand}
   \end{tikzpicture}}
 = - \frac{g^2}{2 m^2} \int \frac{\diff^4 q}{(2\pi)^4}
 \left( \frac{\bk^2}{4} - \bq^2 \right)^2
 \rmi G \left( \frac{k}{2} + q \right) 
  \rmi G \left( \frac{k}{2} - q \right), 
\end{equation}
with $k = (0,\bk)$ and $q = (q^0,\bq)$. 
Substituting the propagator, we obtain 
\begin{equation}
 \tilV (\bk) 
  = \frac{\rmi g^2}{2 m^2} \int \frac{\diff^4 q}{(2\pi)^4} 
  \frac{(\frac{\bk^2}{4} - \bq^2)^2}{[(q^0)^2 - E_+^2 + \rmi \epsilon ] [(q^0)^2 - E_-^2 + \rmi \epsilon ]},
  \label{eq:potential-integral-T=0}
\end{equation}
where we introduced $E_{\pm} \equiv E_{\frac{\bk}{2} \pm q}$.

We can evaluate the integral in Eq.~\eqref{eq:potential-integral-T=0} using the standard field theoretical technique~\cite{Peskin:1995ev}.
Recalling the Feynman parameterization 
\begin{equation}
 \frac{1}{AB} = \int_0^1 \diff x \frac{1}{[x A + (1-x) B]^2},
 \label{eq:feynman-param}
\end{equation}
we first rewrite the potential as 
\begin{equation}
 \tilV (\bk) 
  = \frac{\rmi g^2}{2 m^2} \int^{1}_{0}\diff x\int \frac{\diff^4 q}{(2\pi)^4} 
  \frac{(\frac{\bk^2}{4} - \bq^2)^2}{[(q^0)^2 - x E_+^2 - (1-x) E_-^2 + \rmi \epsilon]^2}.
\end{equation}
The integrand has two poles at 
$q^0 = \sqrt{x E_+^2 + (1-x) E_-^2} - \rmi \epsilon$ 
and $q^0 = - \sqrt{x E_+^2 + (1-x) E_-^2} + \rmi \epsilon$.
Performing the $q^0$-integration by closing the contour in the lower half-plane, we obtain 
\begin{equation}
 \begin{split}
  \tilV (\bk)
  &= - \frac{g^2}{8 m^2} \int_0^1 \diff x 
  \int \frac{\diff^3 q}{(2\pi)^3} 
  \frac{(\frac{\bk^2}{4} - \bq^2)^2}{[x E_+^2 + (1-x) E_-^2]^{3/2}}
  \\
  &= - \frac{g^2}{8 m^2 c_s^3} \int_0^1 \diff x 
   \int \frac{\diff^3 q}{(2\pi)^3} 
   \frac{\bq^4 + (1- 2x)^2 (\bq\cdot \bk)^2-2x(1-x)\bq^2\bk^2 + x^2 (1-x)^2 \bk^4}{[ \bq^2 + x (1-x)\bk^2]^{3/2}}.
 \end{split}
 \label{eq:divergent-V-T=0}
\end{equation}
To obtain the second line, we substituted $E_{\bp} = c_s |\bp|$ \and changed the integration variable from $\bq$ to $\bq^{\prime} = \bq - \frac{1-2x}{2} \bk$.

To regularize the UV divergence in Eq.~\eqref{eq:divergent-V-T=0}, we use the dimensional regularization by changing the spatial dimension from $3$ to $d$~\cite{Peskin:1995ev}.
Introducing the renormalization scale $\lambda$, which has the dimension of the  momentum, and using the rotational invariance and $\diff^d q = q^{d-1} \diff q \diff \Omega_d$
with $\int \diff \Omega_d = \frac{2\pi^{d/2}}{\Gamma (\frac{d}{2})}$, 
we obtain 
\begin{equation}
 \begin{split}
  \tilV (\bk)
  &= - \frac{g^2}{8 m^2 c_s^3} \frac{2 \pi^{d/2}}{(2\pi)^d \Gamma (\frac{d}{2})}
  \lambda^{3-d}
  \int_0^1 \diff x 
  \int_0^\infty \diff q  
  \frac{q^{d+3} + \frac{1}{d} [1- 2(d+2)x + 2(d+2) x^2] \bk^2 q^{d+1} + x^2 (1-x)^2 \bk^4 q^{d-1}}{[ q^2 + x (1-x) \bk^2]^{3/2}}.
 \end{split}
\end{equation}
With the help of the formula
\begin{equation}
 \int_0^\infty \diff q \frac{q^{n-1}}{(q^2 + s)^\alpha}
  = \frac{1}{2} s^{\frac{n}{2} - \alpha} 
  \frac{\Gamma (\frac{n}{2}) \Gamma (\alpha - \frac{n}{2})}{\Gamma (\alpha)},
  \label{eq:regularization}
\end{equation}
we can perform the $q$-integration. 
We then set $d = 3 - \epsilon$ and expand the result with respect to $\epsilon$.
After performing the $x$-integration, we eventually obtain the following result 
\begin{equation}
 \tilV (\bk)
  = \frac{g^2 \bk^4}{32 \pi^2 m^2 c_s^3}
  \left[
   \frac{43}{240} \log \frac{\bk^2}{\lambda^2}  
   - \frac{8261}{1440}
   - \frac{43}{240} \left( \frac{2}{\epsilon} + \log 4 \pi - \gamma + 1 \right)
  \right],
\end{equation}
where $\gamma = 0.57721 \ldots$ denotes Euler's constant.
In the modified minimal subtraction ($\overline{\mathrm{MS}}$) scheme, 
we subtract the combination appearing in the last term, which leads to Eq.~\eqref{eq:V2} in the main text.

\section{Evaluation of two-phonon exchange potential at $T \neq 0$}
\label{sec:two-loop-evaluation-finite-T}

In the finite-temperature case, we need to use the Matsubara Green's function 
\begin{equation}
 \Delta (\rmi \omega_n, \bk) = \frac{1}{\omega_n^2 + E_{\bk}^2}
  \with 
  \omega_n \equiv 2 \pi n T \quad (n \in \Zbb),
\end{equation}
where $T$ denotes the temperature of the superfluid medium.
Then, computing the same two-phonon exchange,
the impurity potential at $T \neq 0$ is given by
\begin{align}
 \tilV_{T} (\bk) 
 =& -\frac{g^2}{2 m^2} 
 T \sum_{l= -\infty}^{\infty}
 \int \frac{\diff^3 q}{(2\pi)^3} 
 \left( \frac{\bk^2}{4} - \bq^2 \right)^2
 \Delta \left( \rmi \omega_l, \frac{\bk}{2} + \bq \right)
 \Delta \left( - \rmi \omega_l, \frac{\bk}{2} - \bq \right).
 \label{eq:potential-finite-T-suppl}
\end{align}
We here use the formula~(see, e.g., Ref.~\cite{LeBellac2000})
\begin{equation}
 \begin{split}
  T \sum_{l= -\infty}^{\infty}
  \Delta \left( \rmi \omega_l, \frac{\bk}{2} + \bq \right)
  \Delta \left( - \rmi \omega_l, \frac{\bk}{2} - \bq \right)
  &= \frac{1}{2 E_+ E_-}
  \left[ 
  \frac{1 + f (E_+) + f (E_-) }{E_+ + E_-}
  - \frac{f (E_+) - f (E_-) }{E_+ - E_-}
  \right],
 \end{split}
\end{equation}
which enables us to rewrite Eq.~\eqref{eq:potential-finite-T-suppl} as
\begin{align}
 \tilV_{T} (\bk) 
 =& - \frac{g^2}{2 m^2} 
 \int \frac{\diff^3 q}{(2\pi)^3} 
 \left( \frac{\bk^2}{4} - \bq^2 \right)^2
 \frac{1}{2 E_+ E_-}
 \left[ 
 \frac{1}{E_+ + E_-}
 + \frac{f (E_+) + f (E_-) }{E_+ + E_-}
 - \frac{f (E_+) - f (E_-) }{E_+ - E_-}
 \right].
\end{align}
where we introduced the Bose distribution 
$f (E) \equiv 1/(\rme^{\beta E} - 1)$ with the inverse temperature 
$\beta \equiv 1/T$.
The first term coincides with the result at $T=0$, obtained after performing the $q^0$-integration of Eq.~\eqref{eq:potential-integral-T=0}. 
We, thus, identify the finite-temperature correction, 
$\tilV_T (\bk) = \tilV (\bk) + \Delta \tilV_T (\bk) $, with 
\begin{equation}
 \Delta \tilV_T (\bk) 
  = - \frac{g^2}{2 m^2} 
 \int \frac{\diff^3 q}{(2\pi)^3} 
 \left( \frac{\bk^2}{4} - \bq^2 \right)^2
 \frac{1}{2 E_+ E_-}
 \left[ 
  \frac{f (E_+) + f (E_-) }{E_+ + E_-}
  - \frac{f (E_+) - f (E_-) }{E_+ - E_-}
 \right].
\end{equation}

After performing the angular integral, we obtain 
\begin{equation}
 \begin{split}
  \Delta \tilV_T (\bk) 
  &= \frac{g^2}{32 \pi^2 m^2 c_s^3} 
  \frac{T^4}{c_s^4} \frac{1}{\pk}
  \int_0^\infty \diff x \frac{1}{\rme^{x} - 1} 
  \Big[
  4 \pk^3 x - 16 \pk x^3 + (\pk^2 - 2 x^2)^2 
  \big[ 
  \log |\pk - 2 x| - \log |\pk + 2 x|
  \big]
  \Big]
  \\
  &= \frac{g^2}{32 \pi^2 m^2 c_s^3} 
  \frac{T^4}{c_s^4} 
  \left[
  \frac{2\pi^2}{3} \pk^2 - \frac{16 \pi^4}{15} 
  + \frac{1}{\pk}
  \int_0^\infty \diff x \frac{1}{\rme^{x} - 1} 
  (\pk^2 - 2 x^2)^2
  \big[ 
  \log |\pk - 2 x| - \log |\pk + 2 x|
  \big]
  \right],
 \end{split}
 \label{eq:tilVT-exact}
\end{equation}
where we introduced $\pk \equiv c_s k/T$.
Equation \eqref{eq:tilVT-exact} gives the finite-temperature correction applicable to 
both intermediate distance $\xi \ll r\ll c_s/T$ and longer distance $r\gg c_s/T$ regimes.

\subsubsection{At intermediate distance $\xi \ll r\ll c_s/T$}

Dividing the integration as $\int^{\infty}_{0}\diff x=\int^{\pk/2}_{0}\diff x+\int^{\infty}_{\pk/2}\diff x$, we can find an analytic expression at intermediate distance $\xi \ll r\ll c_s/T$, or the low-temperature limit.
In the low-temperature limit $T\to 0$, i.e., $\pk\to\infty$ at fixed $k$, 
we can approximate the integral by picking up the contribution from the small-$x$ integral. 
Then, expanding the logarithmic function with 
respect to $\pk^{-1}$, we obtain
\begin{equation}
 \begin{split}
  \frac{1}{\pk}
  \int_0^{\pk/2} \diff x \frac{1}{\rme^{x} - 1}
  (\pk^2 - 2 x^2)^2
  \big[ 
  \log (\pk - 2 x) - \log (\pk + 2 x)
  \big] 
  \simeq - \frac{2\pi^2}{3} \pk^2 + \frac{32 \pi^4}{45} 
  - \frac{128 \pi^6}{135} \frac{1}{\pk^2} + O(\pk^{-4}),
 \end{split}
\end{equation}
where we used an approximation extending the upper bound of the integral to infinity.
The large-$x$ integral and the error associated with the above approximation are suppressed by the exponential factor $\rme^{-\pk/2}$ resulting from the Bose distribution $1/(\rme^x-1)$, which we can safely neglect in the low-temperature limit (see also Fig.~\ref{fig:exact-expansions}).
Substituting this result, we obtain the low-temperature correction to the impurity potential as
\begin{equation}
 \Delta \tilV_T (\bk) 
  = - \frac{g^2}{32 \pi^2 m^2 c_s^3} 
  \frac{T^4}{c_s^4} 
  \left[
   \frac{16 \pi^4}{45} + \frac{128 \pi^6}{135} 
   \frac{T^2}{c_s^2} \frac{1}{\bk^2}
  + O \big( T^4/\bk^4 \big)
  \right].
 \label{eq:tilVT-low}
\end{equation}
Omitting the constant term, which does not contribute at long distance, we obtain Eq.~\eqref{eq:kspace-potential-T} in the main text.
As shown in Fig.~\ref{fig:exact-expansions}, Eq.~\eqref{eq:tilVT-low} agrees with 
the result \eqref{eq:tilVT-exact} with a numerical integration at large $\pk$.

\subsubsection{At longer distance $r\gg c_s/T$}
At a longer distance, there is a simple way to evaluate the leading part of $\tilV_T(\bk)$ by taking only the Matsubara zero mode ($\omega_{l} = 0$) contribution in Eq.~\eqref{eq:potential-finite-T-suppl} as
\begin{align}
 \tilV_{T} (\bk) 
 \simeq - \frac{g^2}{2 m^2} 
 T
 \int \frac{\diff^3 q}{(2\pi)^3} 
 \left( \frac{\bk^2}{4} - \bq^2 \right)^2
 \Delta \left(0, \frac{\bk}{2} + \bq \right)
 \Delta \left(0, \frac{\bk}{2} - \bq \right).
\end{align}
Using the Feynman parameterization~\eqref{eq:feynman-param}, we can write the potential as
\begin{align}
\begin{split}
\tilV_{T} (\bk)
& = -\frac{g^2}{2m^2}T\int^{1}_{0}\diff x\int \frac{\diff^3 q}{(2\pi)^4} 
  \frac{(\frac{\bk^2}{4} - \bq^2)^2}{[x E_+^2 + (1-x) E_-^2]^2} \\
& = -\frac{g^{2}}{2m^{2}c_s^{4}}T
	\int^{1}_{0} \diff x \int \frac{\diff^3 q}{(2\pi)^3}
	\frac{\bq^{4}+\frac{1}{3}(1-10x+10x^2)\bk^{2}\bq^{2}+x^{2}(1-x)^{2}\bk^{4}}
		{[\bq^{2}+x(1-x)\bk^{2}]^{2}}.
    \label{eq:divergent-V-Tfinite}
\end{split}
\end{align}
To obtain the second line, we substituted $E_{\bp} = c_s |\bp|$, changed the integration variable from $\bq$ to $\bq^{\prime} = \bq - \frac{1-2x}{2} \bk$, and used the rotational invariance as is the case for the zero temperature.
Performing the $\bm{q}$-integration with the help of Eq.~\eqref{eq:regularization}
followed by the $x$-integration, we arrive at Eq.~\eqref{eq:potential-k-T-long} in the main text.

\begin{figure}[t]
 \centering
 \includegraphics[width=0.5\linewidth]{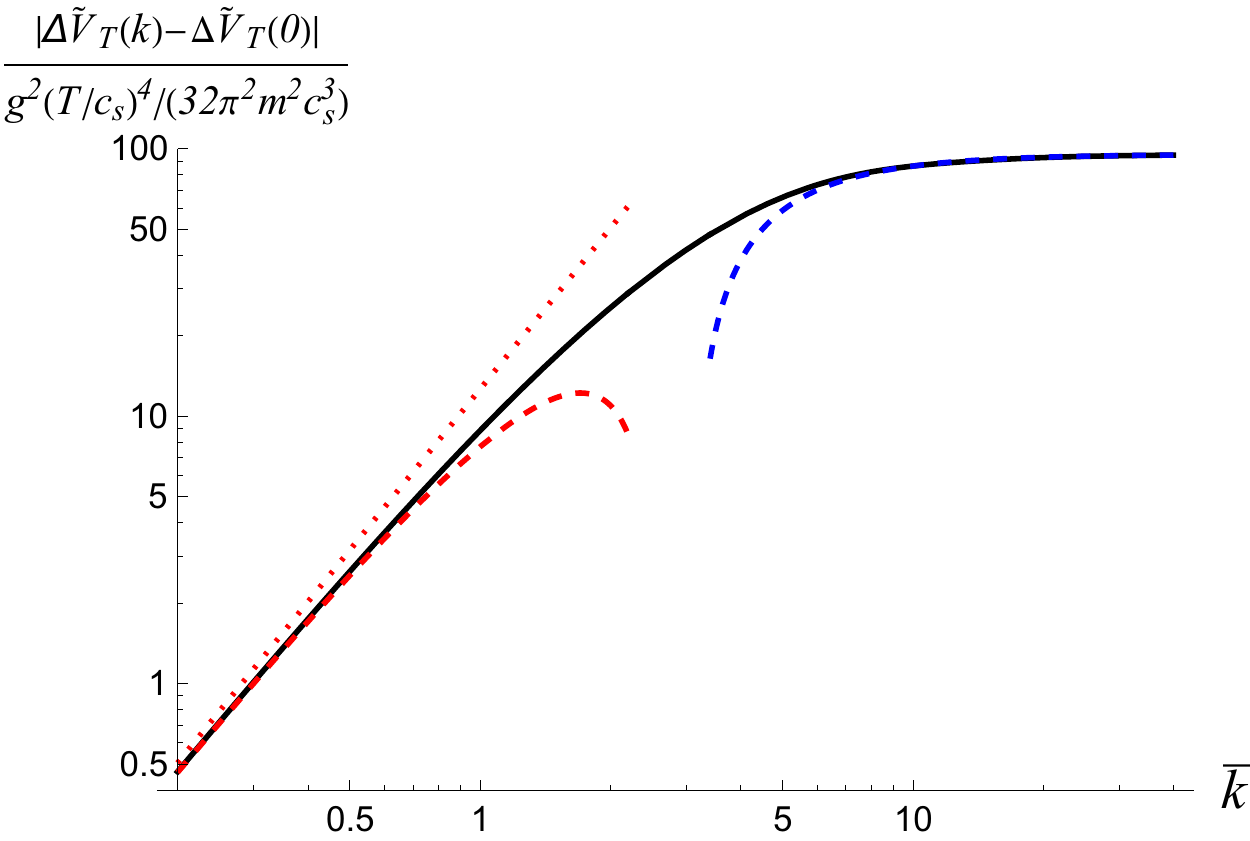}
 \caption{\label{fig:exact-expansions}
Log-log plot of the finite-temperature correction as a function of the dimensionless momentum $\pk = c_s k /T$.
While the solid black curve represents the exact finite-temperature correction given by Eq.~(\ref{eq:tilVT-exact}), the dashed blue and red curves represent the low- and high-temperature correction given by Eqs.~(\ref{eq:tilVT-low}) and~\eqref{eq:tilVT-High}, respectively.
The dotted red curve represents up to the second term in Eq.~\eqref{eq:tilVT-High}, so that the importance of the $\pk^3$ term included in the dashed line is apparent.
 }
\end{figure}

One can also derive this result~\eqref{eq:potential-k-T-long} from the high-temperature limit of Eq.~\eqref{eq:tilVT-exact}. 
In the high-temperature limit $T\to \infty$, i.e., $\pk\to 0$ at fixed $k$, we replace the dimensionless Bose distribution with $1/x$ and expand the result with respect to $\pk$ after integrating over $x$.
Then, we find
\begin{align}
\Delta\tilV_T(\bk)=-\frac{g^2}{32\pi^2m^2c_{s}^{3}}\frac{T^4}{c_s^4}\biggl[\frac{16\pi^4}{15}-\frac{2\pi^2}{3}\pk^2+\frac{\pi^2}{2}\pk^3+O(\pk^4)\biggr],
\label{eq:tilVT-High}
\end{align}
using the cutoff regularization.
This result coincides with Eq.~\eqref{eq:potential-k-T-long} in the main text after omitting the irrelevant constant and $k^2$ terms, which do not contribute at long distance. 
One can also confirm that the approximated result agrees with the finite-temperature correction \eqref{eq:tilVT-exact} with a numerical integration at small $\pk$, as shown in Fig.~\ref{fig:exact-expansions}.

\end{document}